\documentclass[a4paper,twoside]{article}

\usepackage{epsfig}
\usepackage{subfigure}
\usepackage{calc}
\usepackage{amssymb}
\usepackage{amstext}
\usepackage{amsmath}
\usepackage{amsthm}
\usepackage{multicol}
\usepackage{pslatex}
\usepackage{apalike}
\usepackage{cite}
\usepackage{geometry}
\usepackage{SCITEPRESS}     

\begin{document}

\title{Deep Generative Models to Extend Active Directory Graphs with Honeypot Users}

\author{\authorname{Ondřej Lukáš\sup{1}\orcidAuthor{0000-0002-7922-8301}, Sebastian Garcia\sup{1}\orcidAuthor{0000-0001-6238-9910}}
\affiliation{\sup{1}Faculty of Eletrical Engineering , Czech Technical University, Prague, Czech Republic}

\email{lukasond@fel.cvut.cz, sebastian.garcia@agents.fel.cvut.cz}
}


\abstract{
Active Directory (AD) is a crucial element of large organizations, given its central role in managing access to resources. Since AD is used by all users in the organization, it is hard to detect attackers. We propose to generate and place fake users (honeyusers) in AD structures to help detect attacks. However, not any honeyuser will attract attackers. Our method generates honeyusers with a Variational Autoencoder that enriches the AD structure with well-positioned honeyusers. It first learns the embeddings of the original nodes and edges in the AD, then it uses a modified Bidirectional DAG-RNN to encode the parameters of the probability distribution of the latent space of node representations. Finally, it samples nodes from this distribution and uses an MLP to decide where the nodes are connected. The model was evaluated by the similarity of the generated AD with the original, by the positions of the new nodes, by the similarity with GraphRNN and finally by making real intruders attack the generated AD structure to see if they select the honeyusers. Results show that our machine learning model is good enough to generate well-placed honeyusers for existing AD structures so that intruders are lured into them.
}

\onecolumn \maketitle \normalsize \vfill

\section{\uppercase{Introduction}}
\label{sec:introduction}


From the range of attacks that organizations face, those to the internal network are the most critical. Companies such as Sony, Austria Telekom, NTT, and Citrix have been compromised in their internal networks~\cite{sonyhack, austriahack, ntthack, citrixhack}. These attacks are usually to their Active Directory (AD) to gain access to internal resources~\cite{activedirectorymainstream}. AD stores sensitive data, and since it is used by all internal users, it is difficult to detect attacks in the AD by differentiting between normal and attacker behaviors.

There are three common defenses in AD. First, to stop attackers from \textit{accessing} the AD by using network segmentation, by limiting access~\cite{metcalf2015addefense}, by hardening AD configurations, or by monitoring system events~\cite{nurfauzi2020active, metcalf2015addefense}. Second, to detect anomalies in the use of AD\cite{atakillchain}. Third, to use honeyusers. 

A honeyuser is a fake user disguised as a real user and designed to attract attackers~\cite{honeytoken}. Since users should not interact with honeyusers, \textit{any interaction} triggers a detection. Honeyusers have been used for fake bank accounts and database, but rarely in AD. To maximize the chance of being attacked, the correct placement of the honeyuser in the AD is essential. 

We propose a deep learning variational autoencoder model which generates both features and placement location of honeyusers in AD graphs. First, a graph representation of an existing AD is extracted. Second, the graph is encoded using a Bidirectional Directed Acyclic Graph Recurrent Neural Network (DAG-RNN). The latent space of the encoded graphs is represented by a multivariate Normal didstribution. Third, new nodes are sampled from the probability distribution and a Multilayer Perceptron (MLP) is used to predict their position in the extended graph. The model outputs a set of nodes to add to the AD, their features and \textit{where} (to which nodes) they should be connected.

Since AD data is difficult to obtain, we generated syntetic graphs by \textit{boosting} them with a small sample of real AD structures. These syntetic datasets were used to train and evaluate our model against the GraphRNN technique~\cite{you2018graphrnn}. We also evaluated the quality of the honeyusers by publishing a game to attack a real AD on the Internet. This game helped understand if real attackers are more lured into the honeyusers placed by our model.

Results show that the DAG-RNN model can generate new honeyuser-enriched AD graphs that are in average 80\% similar to the original graph. It can also place honeyusers in \textit{organic} positions 94\% of the time. Preliminary results from the real-life game are inconclusive but suggest an attackers' tendency to prefer the DAG-RNN generated honeyusers. 

The contributions of this paper are:
\begin{itemize}
    \item A DAG-RRN autoencoder for extending AD graphs with honeyusers.
    \item The first Bidirectional DAG-RNN models applied to the domain of honeyusers generation.
    \item An evaluation with real-life attackers.
    \item A public implementation of the DAG-RNN model that only depends on Tensorflow 2.
    \item A sythetic dateset of AD graphs.

\end{itemize}
The rest of the paper is organized as follows: Section~\ref{sec:related_work} describes the related work; Section~\ref{sec:Dataset} describes the generation of the dataset; Section~\ref{sec:method} describes the deep learning method; Section~\ref{sec:experiments} describes the evaluations of the model; Section~\ref{sec:results} shows the results; and Section~\ref{sec:Conclusion} makes the conclusion.

\section{\uppercase{Related Work}}
\label{sec:related_work}
Active Directory (AD) has been analyzed as a target due to its importance inside companies~\cite{ADattack-casestudy}, with the most common detection approach being to search the AD logs for anomalies~\cite{AD-ML-AD}.

Common protecting AD solutions include hardening and monitoring tools~\cite{honeypots4windows}, with the main tool for detecting malicious activities being the Advanced Threat Analytics by Microsoft~\cite{advanced-threat-analytics}, which detects abnormal activity. Some tools manage fake accounts~\cite{blue-hive}, but do not generate new honeyusers. The DCEPT tool~\cite{Dcepttool} creates fake accounts in memory of end-points. To our knowledge, there is no research to automatically generate  honeypots in AD~\cite{high-inter-remoteHP}.

In other areas, automation and machine learning methods were used to design honeypots. Techniques include state machines to generate scripts~\cite{ScriptGen} for the honeypot honeyd~\cite{honeyd}. Reinforcement Learning has also been used for generating honeypot responses to extend the duration of the attack~\cite{HPConcealWithRL2018}. Game Theory was also used to place honeypots as a two-player interaction game~\cite{tian2019honeypot}.

Graph Neural Networks (GNN) were used for detection, generation, and classification of graphs. A prominent work is GraphRNN~\cite{you2018graphrnn}, where the graph is iteratively created using two recurrent modules, one for nodes and one for graphs. GraphRNN outperforms Graph convolutional neural networks on the generation of undirected graphs. 

Graph Variational Autoencoders were used to generate small undirected graphs in molecule modelling with success~\cite{simonovsky2018graphvae}. The method, however, lacks good scaling and predefines the maximal size of structures. 

Graph Recurrent Attention Networks~\cite{liao2019gran}, showed success in modeling protein data, exceeding both GrapVAE and GraphRNN. The technique combined recurrent GNN with attention layers.

Directed Acyclic Graphs (DAG) were used with custom RNN cells to analyze a DAG structure and produce simplifications of formulas~\cite{kaluza2018neural}. A DAG-to-DAG also learnt the satisfiability of formulas in propositional logic~\cite{amizadeh2019dg_dagrnn}. Both works used the Encoder-Decoder architecture on top of graph recurrent cells. As far as we know, there are no publications using generative models for honeypot generation. 

\section{\uppercase{Artificial Dataset}}
\label{sec:Dataset}
Production AD environments have sensitive Personally Identifiable Information (PII) from users, therfore it is hard to obtain good datasets of real AD structures. 

We solved the issue by obtaining few real AD structures by signing Non-disclosure Agreements (NDAs) and using these samples for \textit{boosting} the generation of artificial datasets. These datasets maintain the same characteristics of the real AD, with the help and verification of security experts.

We created four artificial datasets which differ in the number of nodes and edges. Each one contains a large number of graphs with similar number of nodes. All graphs are valid Directed Acyclic Graphs that follow the restrictions of the real AD, such as which groups have more users.

\subsection{Extracting Active Directory Data}
The structure of real AD has to be extracted to be used in our model. We used the tool Sharphound~\cite{sharphound} for this.

We filter the real ADs to only retain five node types and their edges. The types used in our datasets are: \textit{User}, \textit{Computer}, \textit{Domain}, \textit{OrganizationalUnit} (OU), and \textit{Group}. The number of edges for the individual graphs is sampled from a Gaussian distribution using parameters estimated from real AD structures.

To generate our four artificial datasets, we then used the random DAG generation of the NetworkX library~\cite{networkx}. All generated graphs in each dataset have the same node-to-edge ratio and node type as the real AD structures. Table~\ref{tab:dataset_details} shows the properties of the datasets. The main difference between them is their size.

\begin{table}[h]
    \caption{Artificial datasets with number of graphs, number of nodes, mean amount of vertices and mean amount of edges.}
    \label{tab:dataset_details}
    \centering
    \begin{tabular}{|p{0.07\textwidth}|p{0.05\textwidth}|p{0.06\textwidth}|p{0.05\textwidth}|p{0.07\textwidth}|}
        \hline
        Dataset & graph size & \# samples &Mean $| V |$&Mean $| E |$ \\
        \hline
        AD15 & 15 & 2,000 & 12.51 &  19.02\\
        \hline
        AD50 & 50 & 2,000 & 39.88 & 65.49\\
        \hline
        AD150 & 150 & 2,500 & 115.11 & 192.49\\
        \hline
         AD500 & 500 & 1,000 & 353.36 & 600.17\\
         \hline
    \end{tabular}
\end{table}

We assume that the number of edges to other nodes is an important criterion that influences why an attacker chooses that user. Therefore, the usefulness of a honeyuser node for being a good target is related to how many connections it has and to which nodes. 

Each of the artificial datasets was splitted for training and validation (4/5), and testing (1/5). The testing was not used until the final evaluation. The training/validation sets were shuffled.

\section{\uppercase{Graph Generation Framework}}
\label{sec:method}
\begin{figure}[!ht]
  \vspace{-0.2cm}
  \centering
  {\epsfig{file = 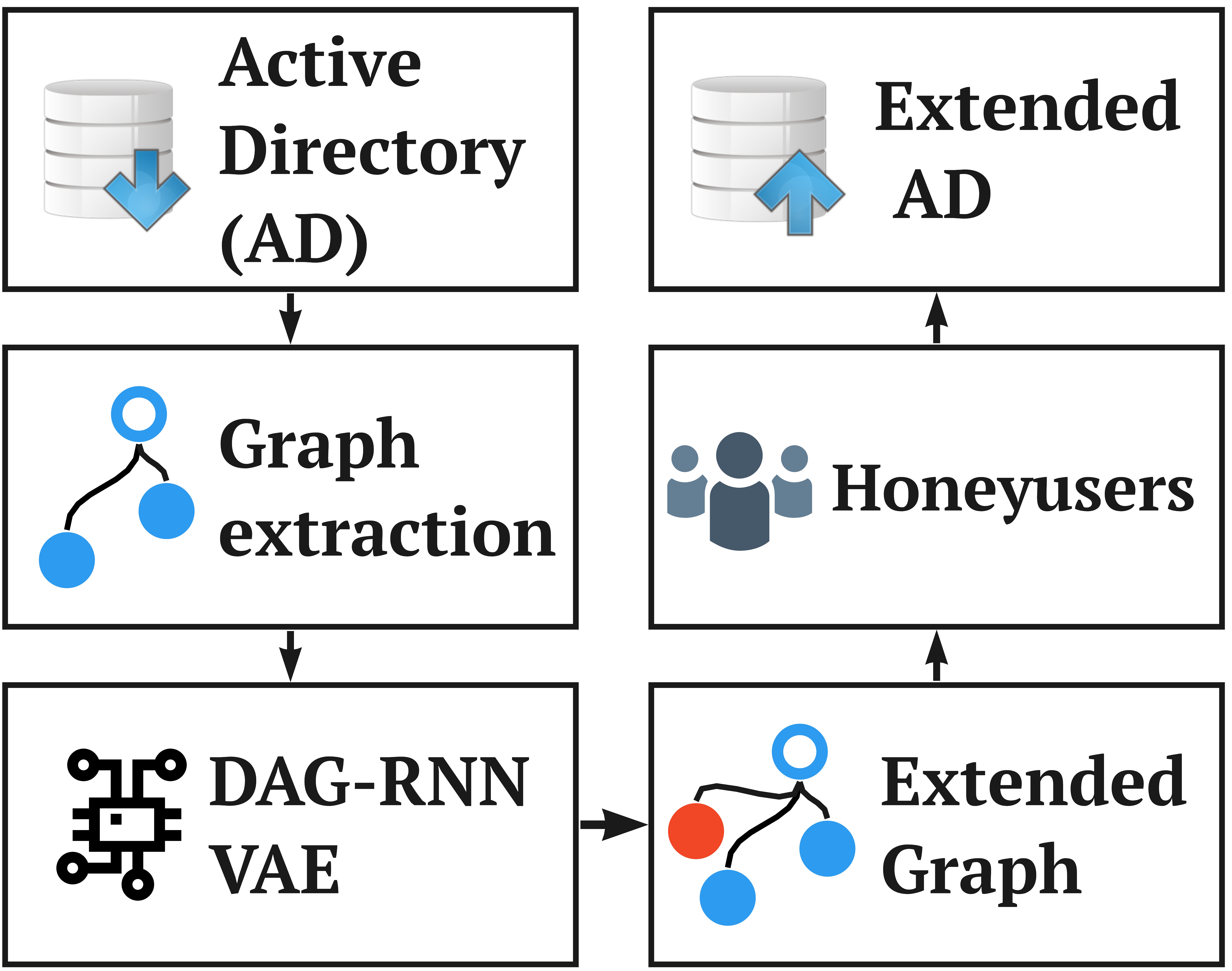, width = 5.5cm}}
  \caption{Diagram of our framework. First, from an AD to graph. Second, embedding of nodes. Third, process nodes with a DAG-RNN Variational Autoencoder. Fourth, predicts locations of nodes. Fifth, enrich the features of nodes. Sixth, inserts the nodes as honeyusers in the AD.}
  \label{fig:framework_pipeline}
 \end{figure}

Our framework starts by creating a graph representation from the AD structure. Then, the graph is encoded into a latent space using the node type embeddings and our bi-directional DAG-RNN encoder. From the encoder, new nodes are sampled and used as input for the decoder, which predicts their placement. Lastly, we generate attributes of the new nodes before inserting them back into the AD. Figure~\ref{fig:framework_pipeline} shows a diagram of the framework.  

\subsection{From AD to Graph Representation}
\label{sec:from-ad-to-input-matrices}
The first step is representing the AD as a directed acyclic graph(DAG). Only six basic node types related to users are present in the graph (Section~\ref{sec:Dataset}). The acyclicity allows for topological sorting of the nodes in the DAG, which is essential for the encoding process. Each graph is represented by $A$, and adjacency node matrix, $A^T$ its transposed version for reverse directionality, and a matrix $X$ that represents one-hot encoded node features. These matrices are zero padded to align the shapes in the mini-batch during the training. The padded nodes are masked during the whole training. The matrix $X$ is input to an embedding layer that outputs the matrix $X'$ with the embeddings that represent similarities between the nodes. 

\subsection{DAG-RNN Variational AutoEncoder}

\begin{figure}[!ht]
  \vspace{-0.2cm}
  \centering
  {\epsfig{file = 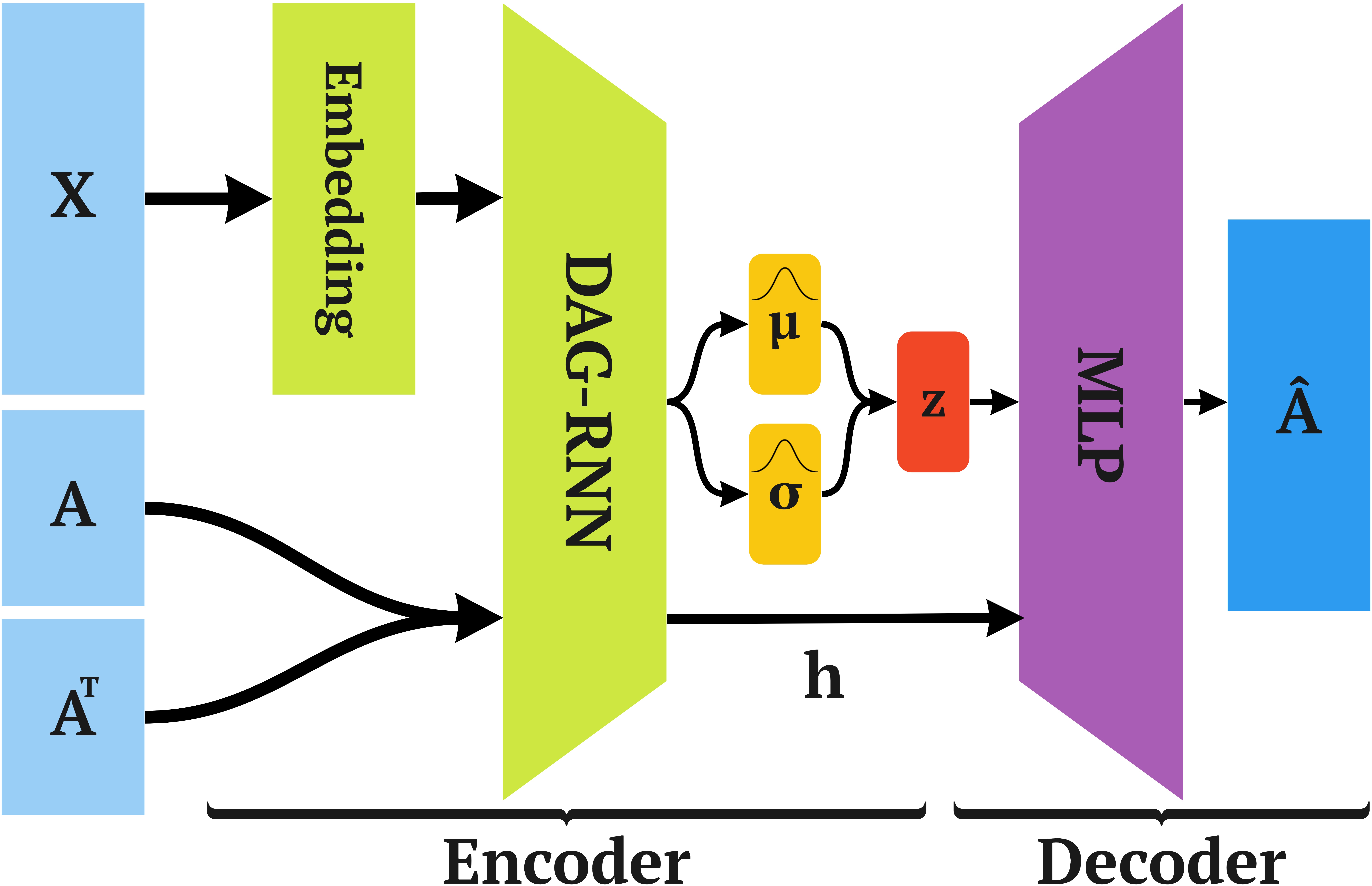, width = 5.5cm}}
  \caption{Overview of the VAE. Inputs are, matrix $X$ (one-hot encoded node types), adjacency matrix $A$, and its transpose $A^T$. Rows follow a topological ordering. The VAE consists of an Encoder (Embedding layer and DAG-RNN layer) which projects the inputs in a latent space $z$, and a Decoder (MLP) which reconstructs the adjacency matrix $\Hat{A}$.}
  \label{fig:DAG_RNN_VAE}
 \end{figure}

The node embeddings and structural information in $A$ and $A^T$ are used in the autoencoding process~\cite{kingma2014autoencoding}. The topology-aware, RNN-based Variational Autoencoder (DAG-RNN VAE) shown in Figure~\ref{fig:DAG_RNN_VAE} learns latent space representation of each node in the graph and can generate new nodes with similar properties.

The DAG-RNN VAE inputs matrices $X'$, $A$ and  $A^T$ and outputs matrix $\Hat{A}$, which contains the placements of the proposed nodes. A multi-variate Gaussian parametrizes the latent space $z$ in which the Encoder represents the original nodes. Such architecture allows sampling of the latent space representation of new nodes. The MLP Decoder predicts the probability of the presence of an edge between a pair of nodes. During the training phase, the model attempts to reconstruct the original adjacency matrix.  During generation, edges \textit{from} the original nodes to newly sampled are predicted.

\subsubsection{DAG-RNN Encoder}
\label{subsub:dag-rnn}
The DAG-RNN layer contains bi-directional Gated Recurrent Units (GRUs), which process the nodes sequentially following the ordering given by $A$ and $A^T$. Unlike a traditional GRU, the output of a  DAG-RNN is not fed back as recurrent input, but stored in the matrix $\overrightarrow{H}$ ($\overleftarrow{H}$ for the unit processing the reversed graph).

In a directed graph there can be multiple previous states (a node can have numerous direct predecessors). Let $G = \langle \mathcal{V}_{G}, \mathcal{E}_{G} \rangle$ be a graph used as input for our method, with $V_G$ the set of vertices of $G$ and $E_G$ the set of edges. By topologically ordering the nodes of $G$ it is guaranteed that $\forall v_i \subseteq V_G$, all of its predecessors have been already been processed in timestamps $t < i$ and their latent space representation is stored in $\overrightarrow{H}$.

When computing the previous state for node $v_i$, we use the corresponding slice of matrix $A$ to create a mask for $\overrightarrow{H}$. With the mask, we can combine the hidden states using summation which results in the previous state for the GRU. Similarly with the reversed graph, we use $A^T$ for masking $\overleftarrow{H}$.

The aggregation of hidden states forces the nodes to be connected in a \textit{similar way} as nodes in the AD. This makes it possible to generate honeyusers that will be part of the most populated groups. 

As a last step, we sum the directional results in $\overrightarrow{H}$ and $\overleftarrow{H}$ using to obtain a single output matrix $H$.

\begin{figure}[!ht]
  \vspace{-0.2cm}
  \centering
  {\epsfig{file = 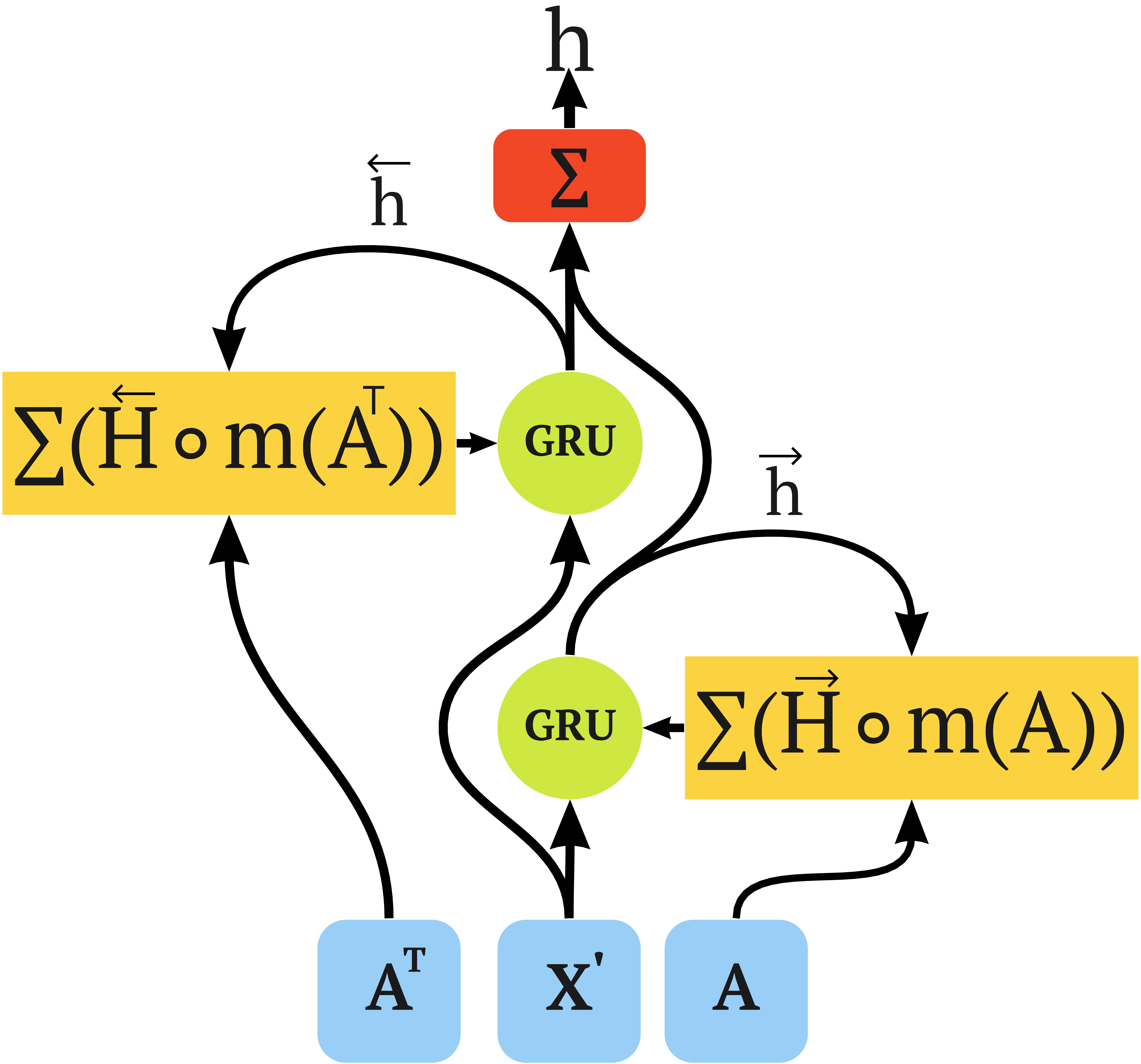, width = 5.5cm}}
  \caption{Bi-directional DAG-RNN layer. The inputs to the GRU cell are the embeddings of the node $X'$ and the aggregation of previous states following the topology of the graph. Matrices $A$ and $A^T$ are used to mask the previous states stored in $\overrightarrow{H}$ and $\overleftarrow{H}$. The outputs of both directions are combined using sum.}
  \label{fig:bidirectiond bal_DAG_RNN}
 \end{figure}

Apart from matrix $H$, which contains the latent space representation of all nodes in the graph, the Encoder outputs vectors $\mu$ and $\sigma$ - the parameters of a multivariate Normal distribution which regularizes the latent space. Each of the parameters is estimated by a single MLP. These parameters are used for (i) latent loss computation, (ii) sampling of the new nodes to be added in the graph. 
\subsubsection{Decoder}\label{subsub:decoder}
The decoder samples one node from the probability distribution for each pre-required honeyuser. The matrix of all requested nodes is $Z$. Then, it pairs the sampled new nodes with the existing nodes doing a Cartesian product between $Z$ and $H$. Each pair is input to an MLP that estimates if the pair should be kept, storing this estimation in $\Hat{A}$. The sigmoid activation of the MLP has a threshold value of 0.2. 
\subsubsection{Loss Functions}\label{subsub:loss}

Since the model is trained all at once, we used a compound weighted loss that is the sum of two functions: a reconstruction loss and a latent loss.

\textbf{The Reconstruction Loss} estimates the auto encoder information loss using the Sigmoid Focal Loss\cite{lin2017focal} (Equation~(\ref{eq:focal_loss}), which is a modification of the binary cross-entropy loss for highly imbalanced classes. In our model, the presence or absence of an edge in $\Hat{A}$ is treated as a binary classification. 

\begin{equation} \label{eq:focal_loss}
    FL(p_t) = -\alpha_{t}(1-p_t)^{\gamma}log(p_t)
\end{equation}

FL is a modification of binary cross-entropy using the parameters $\alpha$ and $\gamma$ to address the imbalance and the different difficulty of classifying classes. The $\gamma$ parameter scales the classification difficulty of the minority class. In FL, ${(1-p_t)}^\gamma$ is a modulating factor, while $pt$ is a notational convenience defined as $p_t = p$ if $y=1$ and $p_t = (1-p)$ otherwise. Where $y$ specifies the ground-truth class, $p$ is the prediction.

\textbf{The Latent Loss} estimates the difference between the distribution of the latent space and the Normal distribution. We used the Kullback-Leibler Divergence~\cite{Kullback-leibler} (Equation~\ref{eq:kl_div}). $D_{KL}$ measures the distance between the latent distribution $Q$ and another distribution (Normal for us) as the prior $P$.

\begin{equation} \label{eq:kl_div}
    D_{KL}\left(P\|Q\right) =  \sum\limits_{x\in X} P(x) log (\frac{P(x)}{Q(x)})
\end{equation}

Where $X$ is the probability space, and $P = N(0,1)$.

The loss function is a weighted sum of the Focal loss and the Latent loss, shown in Equation~\ref{eq:DAGRNNVEA_loss}.

\begin{equation} \label{eq:DAGRNNVEA_loss}
       \mathcal{L} = \frac{n^{2}FL\left(A,\hat{A}\right)}{2} + \lvert z \rvert
       D_{KL}\left(N\left(z_{\mu},z^2_{\sigma}\right)\|N\left(0,1\right)\right) 
\end{equation}

Where $n$ is the number of nodes, $A$ is the adjacency matrix, $\hat{A}$ is the estimated adjacency matrix, and $z_\mu$ and $z_\sigma$ are the estimated parameters of the normal distribution. The Focal Loss is divided by two since we only estimate half of $\hat{A}$, that is a lower triangular matrix.

\subsection{Honeyuser Attributes Generation}
\label{subsec:honeyuser_generation}
For each newly generated node, we still need to generate its AD attributes before adding them to the final graph. For the attributes dependent of the positions, such as \textit{Distinguished Name} (DN), it is necessary to build it based on the position path. For the attributes that are independent of the position, they are randomly generated using external tools such as Faker~\cite{faker}. We verify that the properties of an AD are not violated.

Once the attributes were generated, we insert the extended AD graph back into the original AD server using Powershell cmdlets or LDAP addition queries. 

\subsection{Implementation and Complexity}
The DAG-RNN framework was implemented based on Tensorflow 2 and Keras so it can be used in CUDA GPUs. As far as we know, it is the first complete Tensorflow 2 implementation available.

Sequential processing of the nodes based on the topological ordering results in time complexity $\mathcal{O}(N)$. Since a node $v$ can only be processed after its predecessors have been processed, each pair of nodes $(u,v)$ must processed, which means the memory complexity is quadratic in the size of the input.

\section{\uppercase{Experiments Methodology}}
\label{sec:experiments}

The model was evaluated in three different ways. First, on its ability to encode and reconstruct graph structures. Second, on its ability to extend graphs using the DAG-RNN VAE. Third, on its capacity to generate honeyusers that attract attackers in real-life. 

\subsection{Experimental Setup}
\label{subsec:hyperparameter-training}

The hyperparameters of the model were trained with a mixture of grid search and heuristic expert knowledge. The dimension of the embedding layer is 6. The GRU cell in the encoder consists of 64 units, while the two MLPs that estimate $\mu$ and $\sigma$ have 32 hidden units each.

The MLP encoder has 3 hidden layers with 64, 64 and 32 units respectively, and it uses ReLU activation. The output uses a sigmoid activation.  

The Adam~\cite{kingma2014adam} optimizer is always used for the training with an exponentially decayed learning rate. The initial weights of the DAG-RNN and Decoder MLP are obtained using the Glorot uniform initializer~\cite{Glorot10understandingthe} except for the hidden dense layer for estimating $\sigma$ where the initial weights are 0.

The model was trained in a computer with 32 GB of RAM and an Nvidia Titan V GPU card with 12 GB of RAM. All code is free software~\footnote{https://github.com/stratosphereips/AD-Honeypot}. 

\subsection{Graph Reconstruction}
\label{subsec:structure_reconstruction}

The first evaluation was on graph reconstruction. We measured the generation power to create similar graphs to the original AD by an element-wise comparison of the input embedding adjacency matrix $A$ with the reconstructed matrix $\Hat{A}$. The confusion matrix was created by comparing the same element (position $i,j$) in both matrices: if $A_{i,j}$ and $\Hat{A}_{i,j}$ = 1, it is a TP; if $A_{i,j}$ and $\Hat{A}_{i,j}$ = 0, it is a TN; if $A_{i,j} = 1$ and $\Hat{A}_{i,j} = 0$ it is a FN; if $A_{i,j} = 0$ and $\hat{A}_{i,j} = 1$ it is a FP.

The final metrics used were recall, F1 score, and area under the Precision-Recall Curve (PR AUC).

\subsection{New Nodes Generation}
\label{subsec:generative_ experiments}

The second evaluation was on the generation of new nodes, and used two metrics: Edge Validity Ratio (EVR), and Mean Edge Count Ratio (MECR). They were chosen because in generative models, there is no ground truth to compare with~\cite{GANmeasures}.

\textbf{Edge Validity Ratio} (EVR) is a ratio between the amount of valid edges (possible in an AD) generated for a node and the total amount of generated edges for that node. Equation~\ref{eq:EVR}) shows the EVR, where $\delta^{-}(v)$ is the amount of incoming edges of node $v$ and $\delta^{-}_{valid}(v)$ is the amount of \textit{valid} incoming edges. The final EVR of the graph is the average EVR of all nodes.

\begin{equation}
    EVR(v) = \frac{\delta^{-}_{valid}(v)}{\delta^{-}(v)}
    \label{eq:EVR}
\end{equation}

\textbf{Mean Edge Count Ratio} (MECR) is a ratio between the mean amount of incoming edges in nodes of the original graph ($\delta^{-}_{I}$), and the mean amount of incoming edges in nodes of the generated graph ($\delta^{-}_{G}$). The ratio uses the minimum of these two values as numerator and the maxium as denominator. The mean for the original nodes is defined in Equation~\ref{eq:mean-original} and for the generated nodes in Equation~\ref{eq:mean-generated}.

\begin{equation}
    \delta^{-}_{I} = \frac{1}{|V_{User}|}\sum\limits_{n \in V_{User}} \delta^{-}(n)
\label{eq:mean-original}
\end{equation}

\begin{equation}
    \delta^{-}_{G} = \frac{1}{n}\sum\limits_{i=1}^{n} \delta^{-}(v_{i})
\label{eq:mean-generated}
\end{equation}
The best value of MECR is 1, where the user nodes in the extended graph have in average the same number of incoming edges as the original.

We compare our method with GraphRNN~\cite{you2018graphrnn} on the 2D grid dataset using their proposed Wasserstein distance of node degree distributions between the original and generated nodes. 

\subsection{Evaluation of Nodes as Honeyusers}
\label{subsec:human_experiments}

The third evaluation was on the positions of the nodes as good honeyusers; that is nodes selected by attackers in an AD system. For this we executed a real-life attacking game with two Windows AD systems on the Internet with ~100 users. One AD has the honeyusers placed by our model (edges and features), and the other AD has the same honeyusers but placed in random positions. 

The protocol of the game was as follows: First, users were directed to a webpage where the game was explained~\footnote{https://www.stratosphereips.org/ad-honeypot-game}. Second, one of the two AD is selected randomly and given to the user, where they played by connecting to it with their tools. Third, the user answers three questions with usernames from the AD.

The game used two features from behavioral economic science: first negative rewards (wrong answers decrease the money obtained); second, we donate the final gained money to a charity.  

A question is correct if the selected user is a legitimate domain account, if it is not a honeypot, and if it fulfills the given question.  A task is incorrect if the selected user is a honeyuser or a legitimate domain account that does not fulfill the given question. 

\section{\uppercase{Results \& Discussion}}
\label{sec:results}

\begin{figure}[!ht]
  \vspace{-0.2cm}
  \centering
  {\epsfig{file = 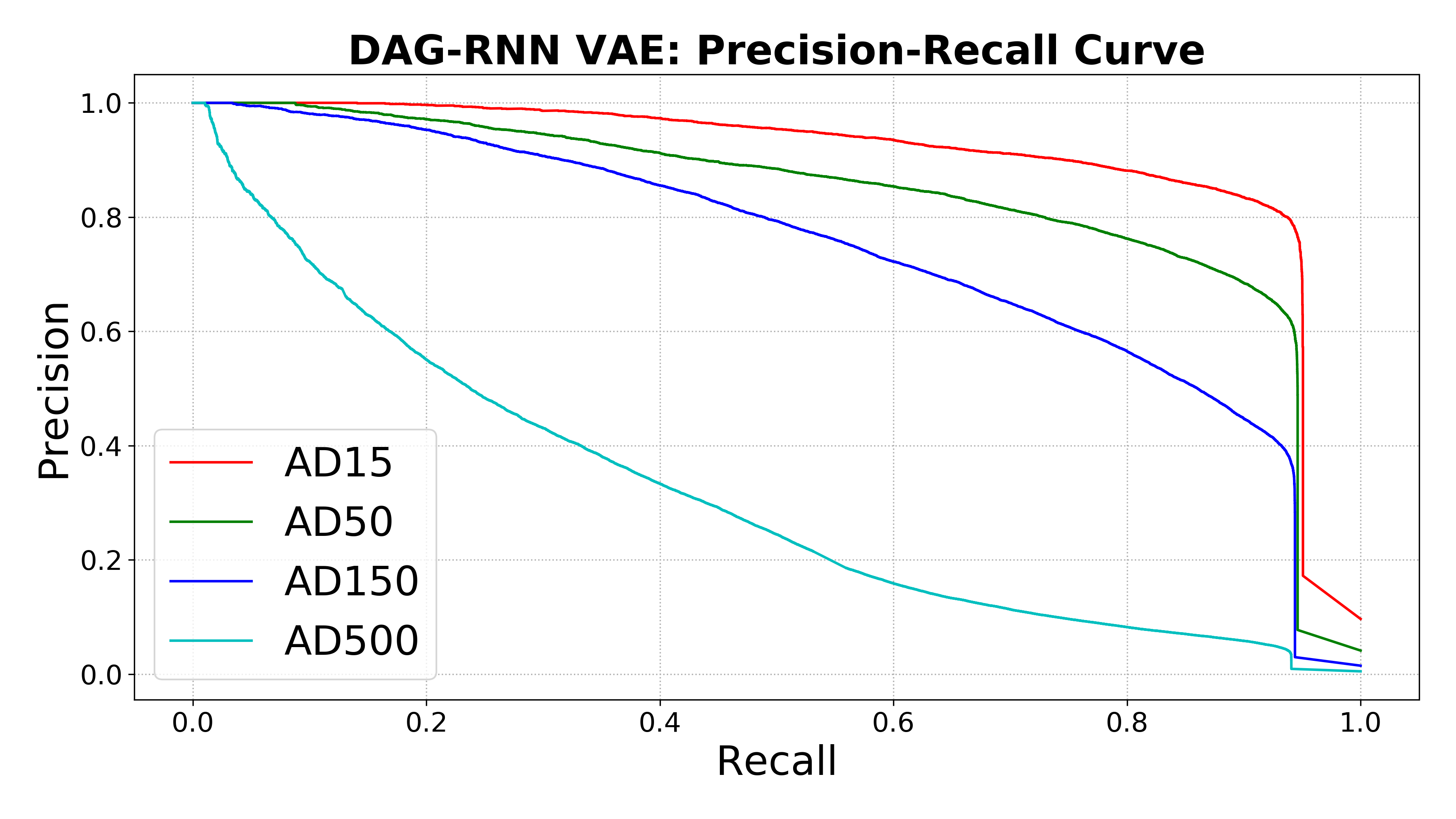, width = 5.5cm}}
  \caption{Graph reconstruction evaluation comparison with Precision-recall curves. Only graphs up to 150 nodes have enough reconstruction precision to be useful.}
  \label{fig:pr_curve_all_dataset}
\end{figure}

\textbf{Results of Graph Reconstruction} There are four sets of results for each of the datasets: AD15, AD50, AD150 and AD500. Table~\ref{table:graph-reconstruction} shows that our model achieves ~80\% precision in the testing set of AD15, AD50, and AD150. However, it only reaches 51\% precision for large datasets of 500 nodes, suggesting that in larger graphs the ability to reconstruct the graph degrades. The F1-score reaches 84\% for middle-size graphs and is close to 60\% for large graphs. Figure~\ref{fig:pr_curve_all_dataset} shows a comparison of precision-recall curves, where it is seen that the largest dataset AD500 has a drop in performance. 

Results suggest that our method can reconstruct graphs with enough precision up to 150 nodes and are useful in the generation of new users, but it struggles with large graphs.


\begin{table}[ht]
\label{tab:amatrix reconstructio}
\caption{Graph reconstruction evaluation metrics.}
\centering
\begin{tabular}{|l|c|c|c|}
  \hline
  Dataset & Precision & Recall &  F1-score \\
  \hline
  AD15 & 80.93\% & 94.5\%6 & 87.22\% \\
  \hline
  AD50 & 79.94\% & 89.48\% & 84.44\% \\
  \hline
  AD150 & 80.38\% & 45.53\% & 58.13\% \\
  \hline
  AD500 & 51.85\% & 72.6\% 7& 60.52\%\\
  \hline
\end{tabular}
\label{table:graph-reconstruction}
\end{table}

\begin{figure}[!ht]
  \vspace{-0.2cm}
  \centering
  {\epsfig{file = 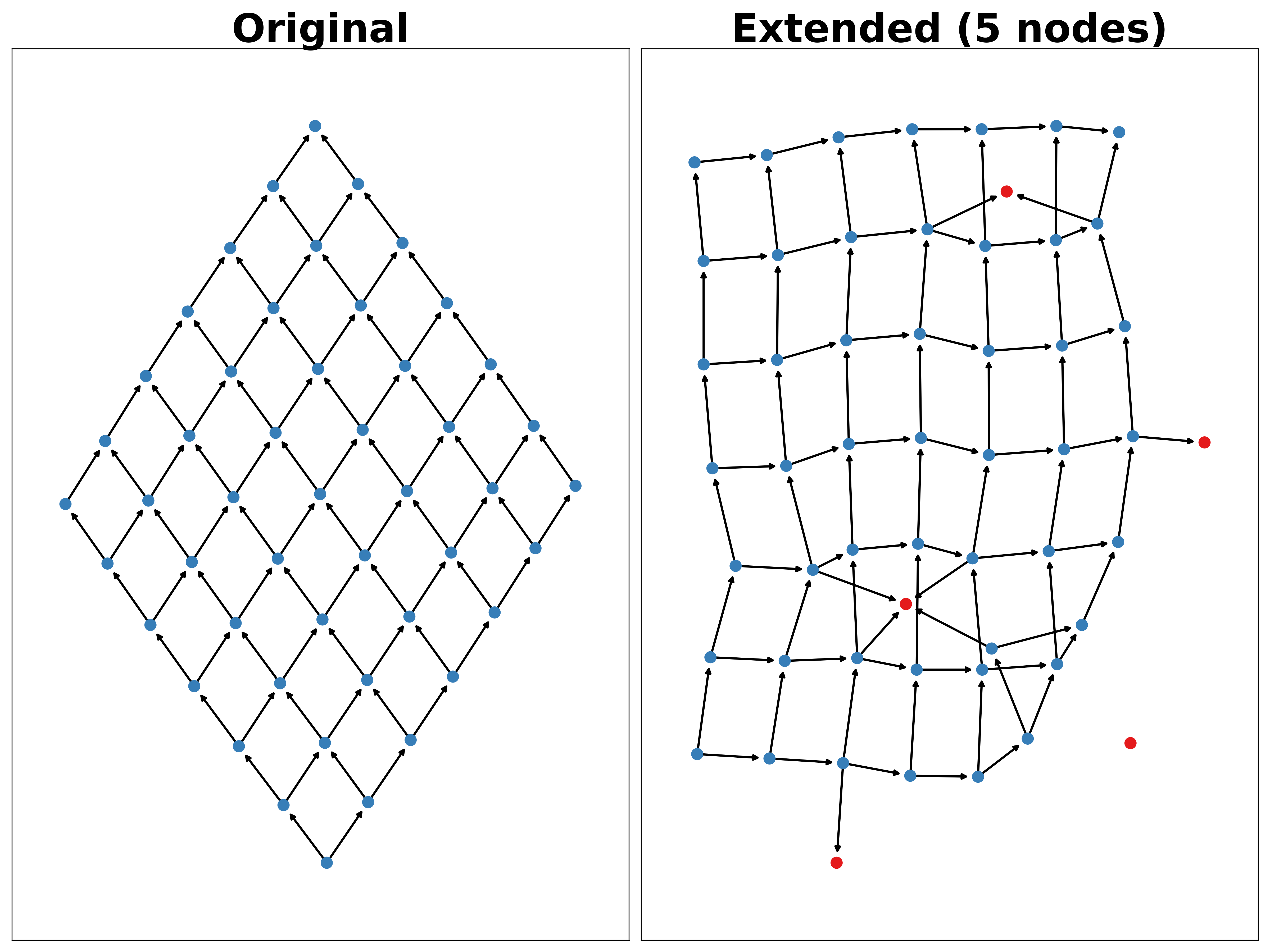, width = 6.5cm}}
  \caption{
  Example of node generation on a 2D grid dataset from GraphRNN. For the DAG-RNN VAE, we added directionality and used nodes of the same type. The Wasserstein distance of node degree was used in the experiment. Our method achieved \textit{0.99} which outperforms all baselines listed in GraphRNN paper. However, the DAG-RNN VAE does not improve the result  of the GraphRNN. The other two metrics used in that paper are not applicable to our method.}
  \label{fig:extented-graphrnn}
\end{figure}

To compare with GraphRNN we trained with 2,000 nodes and evaluated with 500, extending the graphs of size 50 with 5 new nodes. Our average Wasserstein distance is 0.99 (better closer to 0) for \textit{only} the extended nodes, which is better than all the baselines reported in~\cite{you2018graphrnn}. Taking all the nodes into account (original and new) it is 0.15. Figure~\ref{fig:extented-graphrnn} shows an example extended grid.

\textbf{Results of New Nodes Generation} Table~\ref{tab:results-new-nodes-generation} shows the EVR and MECR metrics for each datasets. EVR was better for graphs of 50 nodes, with a precision of 80\% and an F1-score of 84\%. Graphs of 150 and 500 nodes had an F1-score close to 60\%, meaning that for larger graphs we are not generating the same amount of edges or they are connected differently. 

However, with an F1-score ~84\%, we can expect to generate new nodes for graphs of middle size that are \textit{organic} enough to be very similar to the other users of the original graph.

\begin{table}[ht]
\centering
\caption{Node generation evaluation metrics.}
\label{tab:results-new-nodes-generation}
\begin{tabular}{|l|c|c|c|}
  \hline
    Dataset & EVR & MECR \\
  \hline
  AD15 & 68.38\% & 77.53\% \\
  \hline
  AD50 & 72.21\% & 95.42\% \\
  \hline
  AD150 & 69.18\% & 92.42\%\\
  \hline
  AD500 & 58.86\% & 95.23\%\\
  \hline
\end{tabular}
\end{table}

Our worked on graphs up to 150 nodes with good results. This was possible because our specific task of honeyuser generation needed less precision to work.

\begin{figure}[!ht]
  \vspace{-0.2cm}
  \centering
  {\epsfig{file = 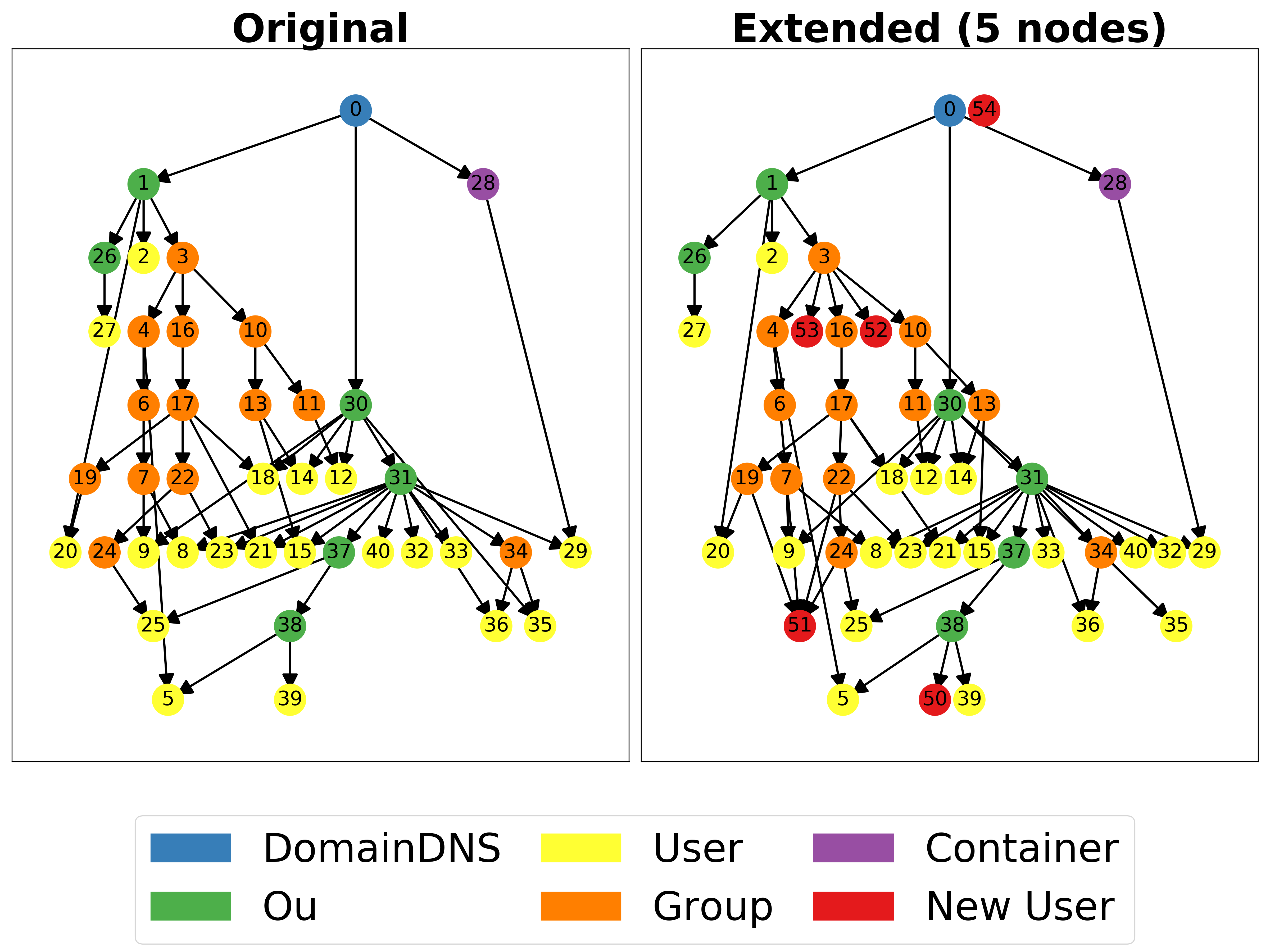, width = 6.5cm}}
  \caption{Example generated graph (right) from an original graph (left) from dataset AD50. There are five user nodes (ids 50-54 in red) inserted. Node 54 is disconnected from the graph and is to be discarded.}
  \label{fig:generated_50}
\end{figure}

Figure~\ref{fig:generated_50} is an example of a generated graph of 50 nodes with honeyusers inserted. Newly added users are depicted in red.

\textbf{Results of Evaluating Nodes as Honeyusers} This result has been hard to measure since it is hard to find real attackers to play the game. With ten participants in the study so far, the results are not statistically significant, but they show trends that we expect to continue for the whole experiment.  

For all three questions, both groups (random AD and generated AD) selected a honeyuser 25\% of the time. At first glance, this may seem to suggest that there is no difference in the generation of edges. However, for the first question participants playing in the \textit{generated AD} selected honeyusers 25\% of the time, compared with the participants in the \textit{random AD} that selected honeyusers 12.5\% of the time. This last result suggests a possible tendency of attackers towards our generated honeyusers.

Given 100 original and 20 honeyusers, the prior probability of choosing a honeyuser was 16.6\%. However, for the first question, the \textit{generated AD} reached 25\% of honeyusers hits, suggesting that honeyusers generated by our model may be selected more than expected. The \textit{random AD} was below this threshold with 12.5\% of honeypot hits.

\section{\uppercase{Conclusion}}
\label{sec:Conclusion}
We presented a deep learning method that ingests Active Directory (AD) structures and generates a similar structure with inserted honeyusers (fake users). The method chooses the position of honeyusers in the AD with a bidirectional topologically sorted DAG-RNN Autoencoder.
The model was evaluated in four ways. First, by generating similar graphs, showing 80\% precision in graphs up to 150 nodes. Second, by placing nodes organically, showing a Mean Edge Count Ratio of 92\%. Third, by comparing with GraphRNN in reconstructing grid graphs and being better than baselines. Fourth, by generating honeyusers that are attractive to attackers in a public real game, showing inconclusive results given the small number of participants, but with preliminary results that seem to suggest that the nodes placed by our RNN are selected slightly more.

The contributions of this work are (i) an application of DAG-RNN in the cybersecurity domain; (ii) a free software implementation of DAG-RNN VAE with GPU acceleration; (iii) a synthetic Active Directory structure dataset; (iv) a framework for real-life AD honeyuser evaluation.

\textbf{Future Work} to improve the experiments with real attackers, to estimate the node type from the embedding, and to include the attractiveness of AD groups.

\section*{\uppercase{Acknowledgements}}
We acknowledge the support of NVIDIA Corporation with the donation of a Titan V GPU for this research. We would also like to thank the Stratosphere team for their support.
\bibliographystyle{apalike}
{\small
\bibliography{references}}
\end{document}